# Chromatic and Spherical Aberration Correction with Hexapole and Quadrupole Fields

Shigeyuki Morishita[1,2,3,4,*], Hidetaka Sawada[1], Norihiro Okoshi[1], Shunsaku Waki[1], Hironori Tanaka[1], Katsunori Ichikawa[1] and Angus Kirkland[3, 4]

[1] *JEOL Ltd., 3-1-2 Musashino, Akishima, Tokyo, 196-8558, Japan*

[2] *JEOL (U.K.) Limited, 1-2 Silver Court, Watchmead, Welwyn Garden City, Hertfordshire, AL7 1LT, U.K.*

[3] *Department of Materials, University of Oxford, Parks Road, Oxford, Oxfordshire, OX1 3PH, U.K.*

[4] *Rosalind Franklin Institute, Harwell Science and Innovation Campus, Didcot, Oxfordshire, OX11 0QX, U.K.*

[*]Corresponding author: shigeyuki.morishita@jeoluk.com

## Abstract

We report the development of a chromatic and spherical aberration corrector based on combinations of hexapole and quadrupole fields. Thick hexapole fields are used to generate negative third order spherical aberration and to correct residual axial and off-axial aberrations. As an alternative to the use of round transfer lenses placed between the hexapoles, a quadrupole multiplet producing superimposed electric and magnetic quadrupole fields is used to produce negative chromatic aberration. This quadrupole multiplet also functions as a transfer doublet within the corrector. The simultaneous correction of chromatic and spherical aberrations using this corrector design is described, and a resolution improvement is demonstrated for cases where the energy spread is limiting.

## 1. Introduction

The correction of the spherical and chromatic aberration inherent in round electromagnetic objective lenses [1] is important for high-resolution imaging using either transmission electron microscopy (TEM) and scanning transmission electron microscopy (STEM). Various optical geometries for spherical aberration correctors have been proposed and experimentally demonstrated and have significantly improved the spatial resolution [2,3] in TEM and STEM [4,5] and the resolution of STEM EELS/EDS mapping [6]. They have notably improved resolution at low accelerating voltages [7] and for samples in magnetic field-free environments [8]. There are two fundamental types of spherical aberration correctors that have been reported based on respectively hexapole elements [2,9] and quadrupole-octupole combinations [3]. In hexapole correctors, thick hexapole fields generate a negative third order spherical aberration from a combination aberration from the primary three-fold astigmatism in each hexapole. In quadrupole-octupole correctors, a negative spherical aberration is generated by using octupole fields acting on elliptical beams formed by quadrupoles. Both types of correctors have been successfully commercialised, and the initial designs have also been refined to also enable correction of higher-order axial and off-axial geometric aberrations [10-12].

In addition to third order spherical aberration ($C_s = C_{3,0}$) and residual axial geometric aberrations, first order chromatic focus ($C_c = C_{c1,1,0}$) and astigmatism ($C_{c1,1,2}$) require correction for high resolution TEM imaging where the partial temporal coherence due to the intrinsic energy width of the incident beam is limiting [13]*. One option for reducing the resolution limiting effect of chromatic aberration is by monochromation of the primary beam [14,15]. However, the beam current of a monochromated beam is inherently reduced and the effect of chromatic aberration due to the energy spread caused by inelastic scattering within thicker samples cannot be compensated using this approach. These disadvantages can be overcome by installing a chromatic aberration corrector to compensate the intrinsic positive chromatic aberrations of the round electromagnetic objective lens. There have been several projects that have reported the simultaneous correction of both geometric and chromatic aberrations, for both bright field phase contrast TEM imaging and high-resolution energy-filtered TEM [16,17]. The influence of chromatic aberrations is more pronounced for low-energy electrons and atomic-resolution TEM imaging of two-dimensional materials has been demonstrated at low accelerating voltages using instruments with chromatic correctors [18, 19]. Chromatic aberration correction is also important for TEM imaging of thick samples for which inelastic electron scattering degrades the image quality which has a particular relevance to high resolution imaging of biological specimens in a cryo state [20,21].

---

*In this paper we adopt the aberration notation due to Krivanek [3] to include chromatic aberrations. In the general case this notation is $C_{ci,\,n,m}$ with the lower-case $c$ indicating a chromatic aberration coefficient and $i$ the power in $(\Delta E/E)^i$. Where $i$ is not specified it is assumed unity. By convention $n$ and $m$ are the radial and azimuthal order of the coefficient. For simplicity in parts of the text we use $C_c$ to denote all first order chromatic aberration coefficients and $C_s$ to denote the coefficient $C_{3,0}$.

All chromatic aberration correctors developed so far consist of combinations of quadrupoles and octupoles [13,22] and are subsequently here referred to as QO correctors. In these designs superimposed electric and magnetic quadrupole fields generate negative $C_{c1,1,0}$ and $C_{c1,1,2}$ and octupole fields generate a negative $C_{3,0}$ on elliptical beams. However, for QO-correctors residual high-order geometric aberrations and off-axial aberrations tend to be larger than those in hexapole correctors [23]. Hence to reduce geometric aberrations, a tandem combination of a hexapole $C_s$ corrector and quadrupole $C_c$ corrector has been previously reported and proof of principle experiments demonstrated [24,25].

In this study, as an alternative to connecting a hexapole $C_s$ corrector and a quadrupole $C_c$ corrector in series, we report the design of a single corrector that corrects both $C_c$ and $C_s$. In the optical arrangement described, it is also possible to superimpose octupole fields on each multipole within the corrector. This combination aberrations of these octupole, hexapole and quadrupole fields can further be used to compensate for the remaining residual axial aberrations.

## 2. Optical Configuration of a Hexapole-Quadrupole (HQ) Corrector

### Optical Concepts

In hexapole $C_s$ correctors, two or more thick multipoles (formed from hexapole or dodecapole elements) are used to generate hexapole fields which are connected by transfer doublets formed from pairs of round lenses [26, 27]. By placing the hexapoles at conjugate planes, the primary three-fold astigmatism created by the hexapole fields is cancelled and the thick hexapole field produces a negative $C_{3,0}$. In this study, we also use thick hexapole fields, but the transfer round lens pairs are replaced by a quadrupole multiplet which acts as a transfer optic [28]. In this arrangement the superposition of electric and magnetic fields in the quadrupole multiplet, allows the correction of $C_{c1,1,0}$ and $C_{c1,1,2}$. In summary for the proposed hexapole-quadrupole (HQ) corrector, $C_{3,0}$ is corrected by thick hexapole fields and $C_{c1,1,0}$ and $C_{c1,1,2}$ are corrected by the transfer quadrupoles.

### Thin-Quadrupole Quadruplet

Transfer doublets play an important role in hexapole correctors by forming conjugate optical planes. Whereas a single lens which can form an amplitude conjugate plane, a transfer doublet can form a plane in which both amplitude and phase are conjugate. Without the inclusion of transfer doublets, combination aberrations from thick hexapoles degrade the image quality in hexapole correctors. In the HQ corrector described here, each conventional round lens doublet is replaced by a quadrupole multiplet. In comparison to a transfer doublet, consisting of two round lenses separated by a specific distance with a controlled focal length, suitable quadrupole multiplets with specific dimensions can also form a conjugate plane. One particularly useful quadrupole multiplet is an antisymmetric quadrupole quadruplet sometimes known as a Russian quadruplet [29,30].

We here outline the formation of a conjugate optical plane using a quadrupole quadruplet and for simplicity, assume that the quadrupoles are thin. For these calculations, a ray transfer matrix analysis [29,30] was used. In this method, a position $r(z)$ and a slope $r'(z)$ at a plane position $z$ are represented as:

$$\begin{pmatrix} r(z) \\ r'(z) \end{pmatrix} = M \begin{pmatrix} r_0 \\ r'_0 \end{pmatrix} \qquad (1)$$

Using a 2 x 2 matrix transfer matrix, $M$, for an optical element, with a position $r_0$, a slope $r_0'$ at an initial plane $z = 0$, the transfer matrix of a drift space with a distance $d$ is:

$$D(d) = \begin{pmatrix} 1 & d \\ 0 & 1 \end{pmatrix}. \qquad (2)$$

Under the assumed thin-quadrupole approximation, the transfer matrices of a quadrupole with a focal length $f$ in the convergent (C) and divergent (D) directions shown in Fig.1 are:

$$Q_{\mathrm{C}}(f) = \begin{pmatrix} 1 & 0 \\ -1/f & 1 \end{pmatrix}, \qquad (3)$$

$$Q_{\mathrm{D}}(f) = \begin{pmatrix} 1 & 0 \\ 1/f & 1 \end{pmatrix}. \qquad (4)$$

Here, we consider an antisymmetric quadrupole quadruplet with quadrupole polarities DCDC in the $x$ direction and CDCD in $y$ direction.

The transfer matrices for this quadrupole quadruplet (Russian quadruplet) in $x$ and $y$ directions can hence be written as:

$$R_x = \begin{pmatrix} x_{1,1} & x_{1,2} \\ x_{2,1} & x_{2,2} \end{pmatrix} = D(d_1) \cdot Q_{\mathrm{C}}(f_1) \cdot D(d_2) \cdot Q_{\mathrm{D}}(f_2) \cdot D(d_3) \cdot Q_{\mathrm{C}}(f_2) \cdot D(d_2) \cdot Q_{\mathrm{D}}(f_1) \cdot D(d_1), \qquad (5)$$

$$R_y = \begin{pmatrix} y_{1,1} & y_{1,2} \\ y_{2,1} & y_{2,2} \end{pmatrix} = D(d_1) \cdot Q_{\mathrm{D}}(f_1) \cdot D(d_2) \cdot Q_{\mathrm{C}}(f_2) \cdot D(d_3) \cdot Q_{\mathrm{D}}(f_2) \cdot D(d_2) \cdot Q_{\mathrm{C}}(f_1) \cdot D(d_1). \qquad (6)$$

Solving for the conditions

$$x_{1,2} = x_{2,1} = 0 \;\text{ and }\; x_{1,1} = y_{1,1}, \qquad (7)$$

and substituting

$$f_1 = f \;\text{ and }\; d_2 = d, \qquad (8)$$

gives the conjugate condition for this quadrupole quadruplet as:

$$d_1 = \frac{f^2}{d} \qquad (9)$$

$$d_3 = \frac{2f^2}{d} \qquad (10)$$

$$f_2 = f \qquad (11)$$

with the relevant transfer matrices simplifying to:

$$R_x = R_y = \begin{pmatrix} -1 & 0 \\ 0 & -1 \end{pmatrix} \qquad (12)$$

The above analysis shows that both the position and slope at a $z_0$ have negative unit magnification at the plane $z_1$, and that they are constant under a 180° rotation operation.

Illustrative ray diagrams for conditions satisfying this solution are shown in Fig. 2.

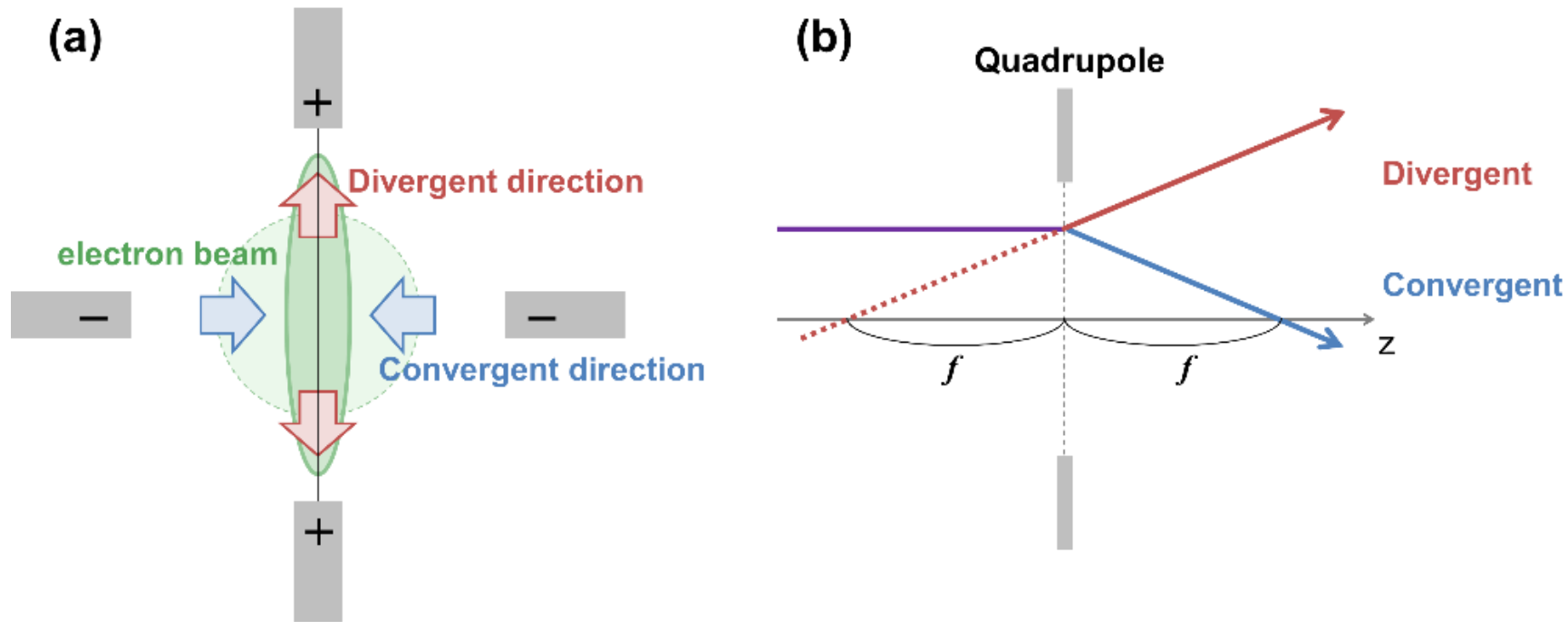

Figure 1: Schematic diagram of a thin quadrupole and electron beam trajectories in (a) the x-y and (b) the x-z planes.

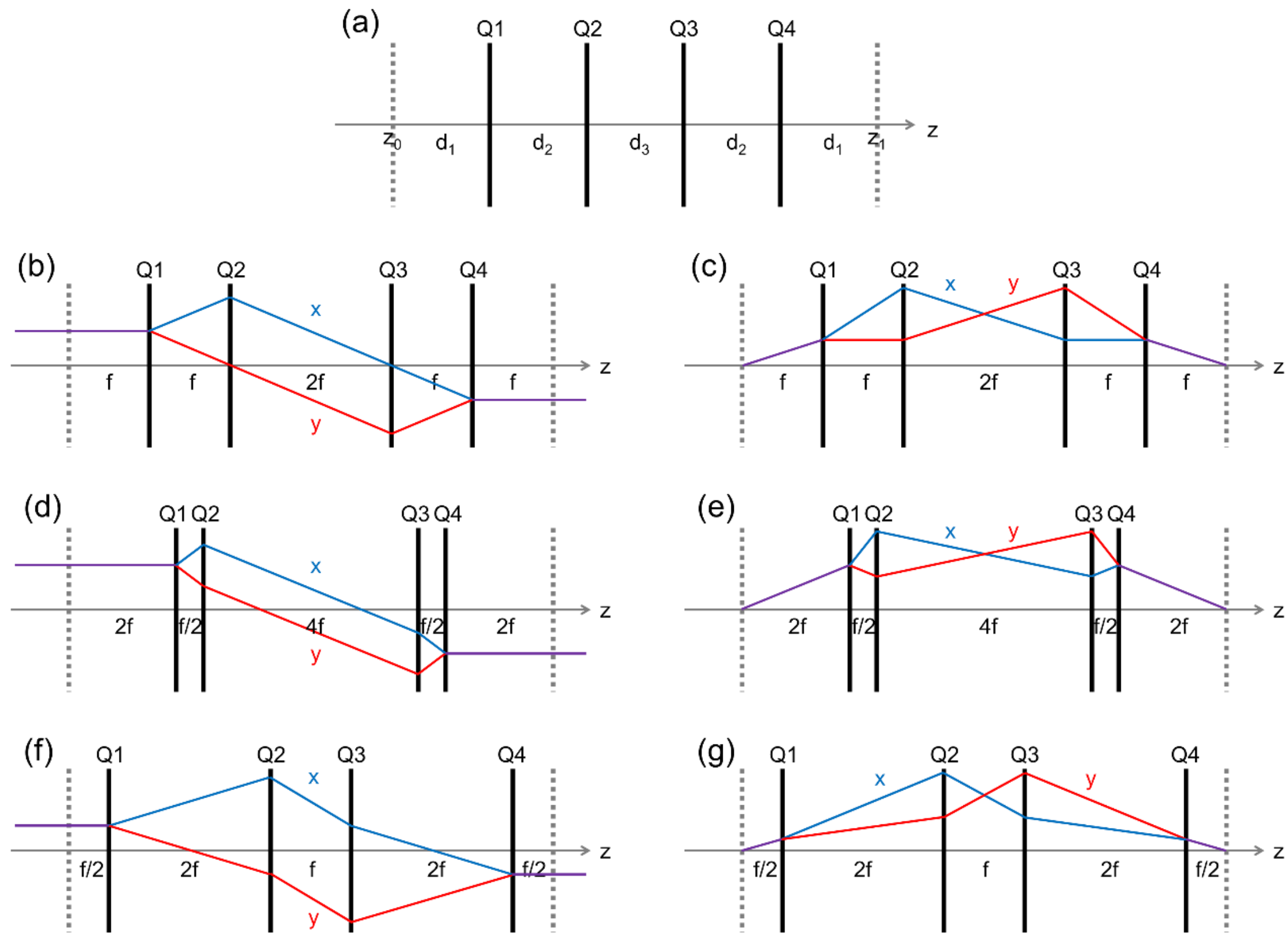

Figure 2: Ray diagrams for thin quadrupole quadruplets under various conditions with a conjugate plane. Typical conditions for $f = d$ are shown in (b) and (c). Other geometries (d, e) $f = 2d$ and (f, g) $f = d/2$ are also shown for comparison.

**Thick-Quadrupole Transfer Multiplets**

For the design of practical $C_c$ correctors, thick quadrupoles instead of thin quadrupoles are required to generate a sufficiently large negative $C_c$. However, any quadrupole multiplet must also form a conjugate plane if it is to be used as an alternative to a round lens transfer doublet. The transfer matrices of a thick quadrupole element for directions C and D respectively become [25]:

$$Q_{\mathrm{C}}(\alpha,t) = \begin{pmatrix} \cos(\alpha t) & \frac{1}{\alpha}\sin(\alpha t) \\ -\alpha\sin(\alpha t) & \cos(\alpha t) \end{pmatrix}, \tag{13}$$

$$Q_{\mathrm{D}}(\alpha,t) = \begin{pmatrix} \cosh(\alpha t) & \frac{1}{\alpha}\sinh(\alpha t) \\ \alpha\sinh(\alpha t) & \cosh(\alpha t) \end{pmatrix}, \tag{14}$$

where $t$ is a thickness of the quadrupole and $\alpha$ is the strength of the quadrupole field given by:

$$\alpha = \sqrt{\pm\left(\frac{2(mc^2+eU)}{2mc^2+eU}\right)\frac{V}{Ub^2} \pm \sqrt{\frac{ec^2}{2mc^2U+eU^2}}\frac{2\mu_0 NI}{b^2}}, \tag{15}$$

where $U$, $b$, $V$ and $NI$ are the accelerating voltage, the bore radius of the quadrupole, the voltages of the electric quadrupole and the ampere turns of the magnetic quadrupole, respectively. The signs are determined by the polarity of each field. In equation (15), $m$, $e$, $c$, and $\mu_0$ represent the electron mass, the absolute value of the electron charge, the speed of light, and the magnetic permeability in vacuum.

The transfer matrices of an antisymmetric quadrupole quadruplet thus become:

$$R_x = D(d_1)\cdot Q_{\mathrm{C}}(\alpha_1,t_1)\cdot D(d_2)\cdot Q_{\mathrm{D}}(\alpha_2,t_2)\cdot D(d_3)\cdot Q_{\mathrm{C}}(\alpha_2,t_2)\cdot D(d_2)\cdot Q_{\mathrm{D}}(\alpha_1,t_1)\cdot D(d_1), \tag{16}$$

$$R_y = D(d_1)\cdot Q_{\mathrm{D}}(\alpha_1,t_1)\cdot D(d_2)\cdot Q_{\mathrm{C}}(\alpha_2,t_2)\cdot D(d_3)\cdot Q_{\mathrm{D}}(\alpha_2,t_2)\cdot D(d_2)\cdot Q_{\mathrm{C}}(\alpha_1,t_1)\cdot D(d_1). \tag{17}$$

The dimensions of the quadrupole quadruplet must also satisfy the following condition;

$$R_x = R_y = \begin{pmatrix} -1 & 0 \\ 0 & -1 \end{pmatrix} . \tag{18}$$

In this case, however, it is difficult to derive analytical solutions and hence the equations describing the transfer matrices were solved numerically. One of the most important parameters of the quadrupole quadruplet in the system described here is $d_1$ which is the distance between the conjugate plane and the entrance plane of the first quadrupole as shown in Figure 3; if $d_1$ is small, the hexapoles cannot be located without physical interference with the outer quadrupoles. Our calculations show that $d_1$ can be controlled by changing the width of $d_2$, $d_3$, $d_4$ and the thickness of quadrupoles $t_1$, $t_2$, $t_3$, $t_4$. Figure 3 shows selected trajectories in thick quadrupole quadruplets. We note here that although a conjugate plane can also be generated by using more than four quadrupoles, we have used the quadrupole quadruplet as the minimum configuration to reduce the overall number of elements in the corrector.

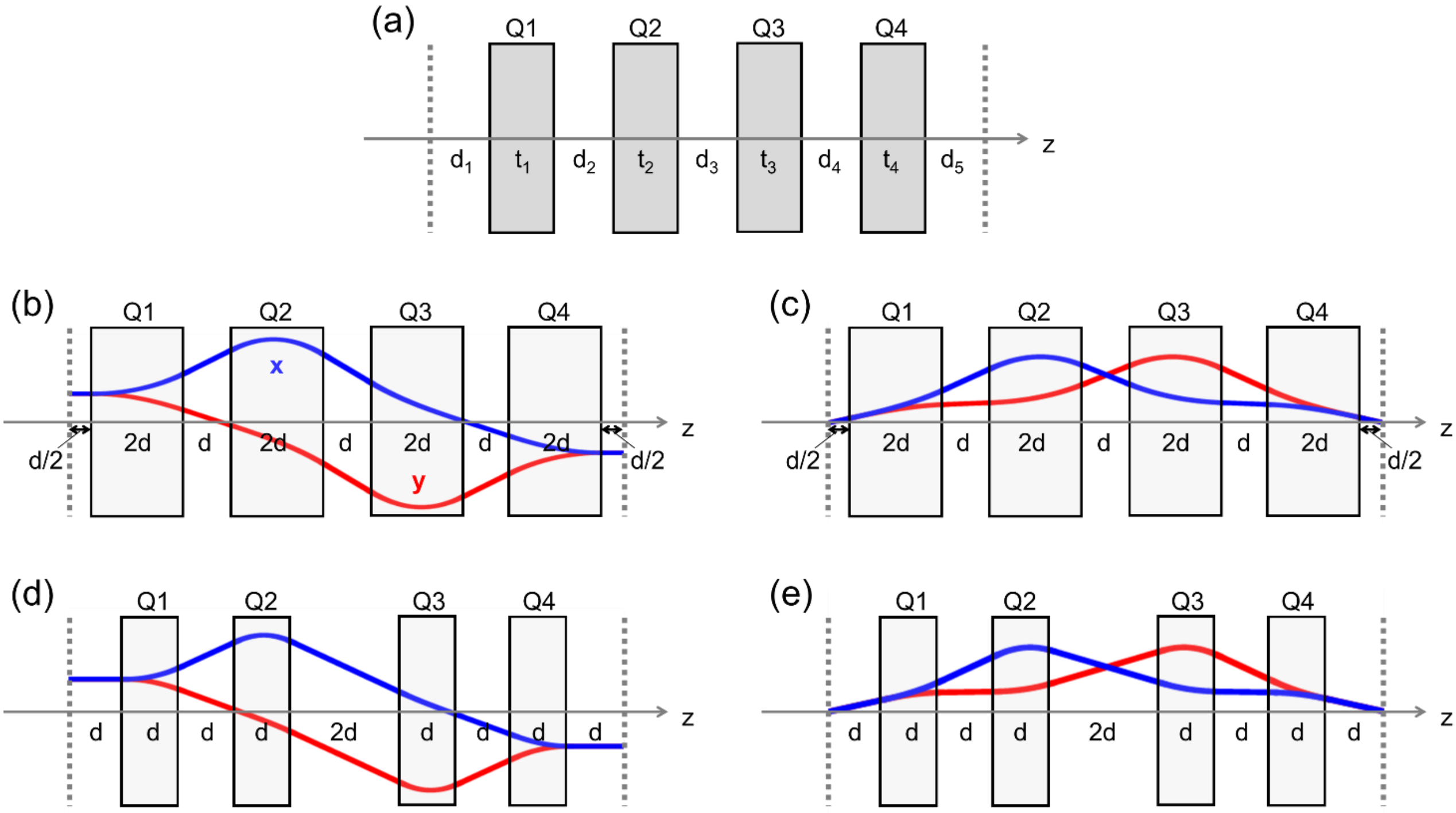

Figure 3: Ray diagrams for thick quadrupole quadruplets under conditions forming a conjugate plane. Typical reduced dimensions are given.

**Overall configuration of the $C_c/C_s$ corrector**

Figure 4 shows the overall configuration of the HQ $C_c/C_s$ corrector. Two hexapole fields (H1, H2) are used to generate a negative $C_{3,0}$ as in conventional hexapole correctors. By connecting the hexapoles with suitable quadrupole multiplets instead of conventional round lens doublets, the two hexapole fields can be placed at conjugate planes. However, in the HQ corrector, additional aberrations are generated due to combination of the hexapole and quadrupole fields, including 2nd order coma, 4th order coma and 5-fold astigmatism. To minimise these parasitic aberrations, quadrupole octuplets consisting of two sets of quadrupole quadruplets with a symmetric trajectory are used for the transfer optics. In addition, a magnetic multipole is also placed at the center conjugate plane of the hexapole fields, to control higher order geometric aberrations.

To correct $C_{c1,1,0}$ and $C_{c1,1,2}$ a superposition of electric and magnetic quadrupole fields is used in the corrector. The electric and magnetic fields are superimposed with opposite polarity and due to the difference in the electron energy dependence of the deflection forces, negative $C_{c1,1,0}$ and $C_{c1,1,2}$ are generated in the inner poles (Q2, Q3, Q6, Q7), where an elliptical (line focus) beam is located. The superposition of electric and magnetic fields in the outer poles (Q1, Q4, Q5, Q8) also increases the negative $C_{c1,1,0}$ and $C_{c1,1,2}$ although the contribution to the overall correction of the outer poles is much smaller than that of inner poles. Hence, it is possible to correct $C_{c1,1,0}$ and $C_{c1,1,2}$ in all directions by generating negative $C_{c1,1,0}$ in the x direction at Q2 and Q7 and in the y direction at Q3 and Q6.

In addition to the primary hexapole or quadrupole magnetic fields, the multipoles are wound with additional elements included to provide further multipole magnetic fields which create dipole, hexapole and octapole fields that can be used for alignment and control of higher order aberrations. Finally, to reduce undesirable combination aberrations between the objective lens and the corrector, a transfer doublet consisting of a pair of round lenses is included in the overall design.

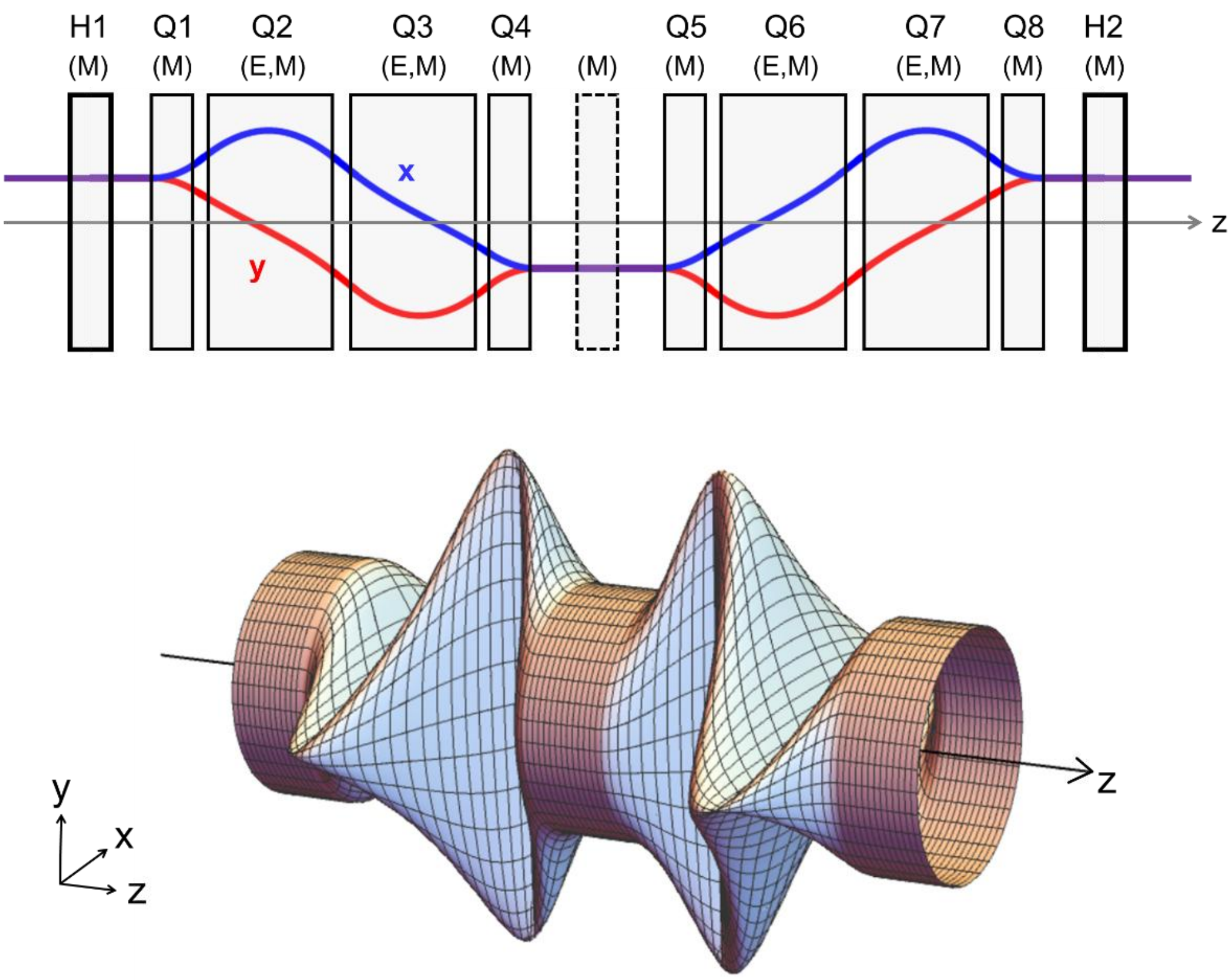

Figure 4: (Upper) Configuration and ray diagram in the x-z and y-z plane of the HQ $C_c/C_s$ corrector. E and M indicate electric and magnetic fields, respectively. (Lower) Three-dimensional first-order trajectories through the corrector.

## 3. Experimental

### Evaluation of Aberrations

The HQ-type $C_c/C_s$ corrector described above is shown in Fig. 5. The height of the corrector is *ca.* 72 cm and the objective lens and corrector are connected using a conventional round lens doublet. In general, the power for each optical element scales with the electron energy and for this corrector, the electric pole voltages required at 200 kV and 300 kV are approximately 2.5 times and 4.6 times higher, than those required for correction at 100 kV. The corrector was designed to operate at a maximum accelerating voltage of 300 kV. However, at lower voltages, individual optical elements have a stronger effect on the electron beam, which facilitates alignment. To demonstrate the optical performance of the HQ corrector, $C_c$ and $C_s$ were corrected, and TEM images were recorded at 200 kV.

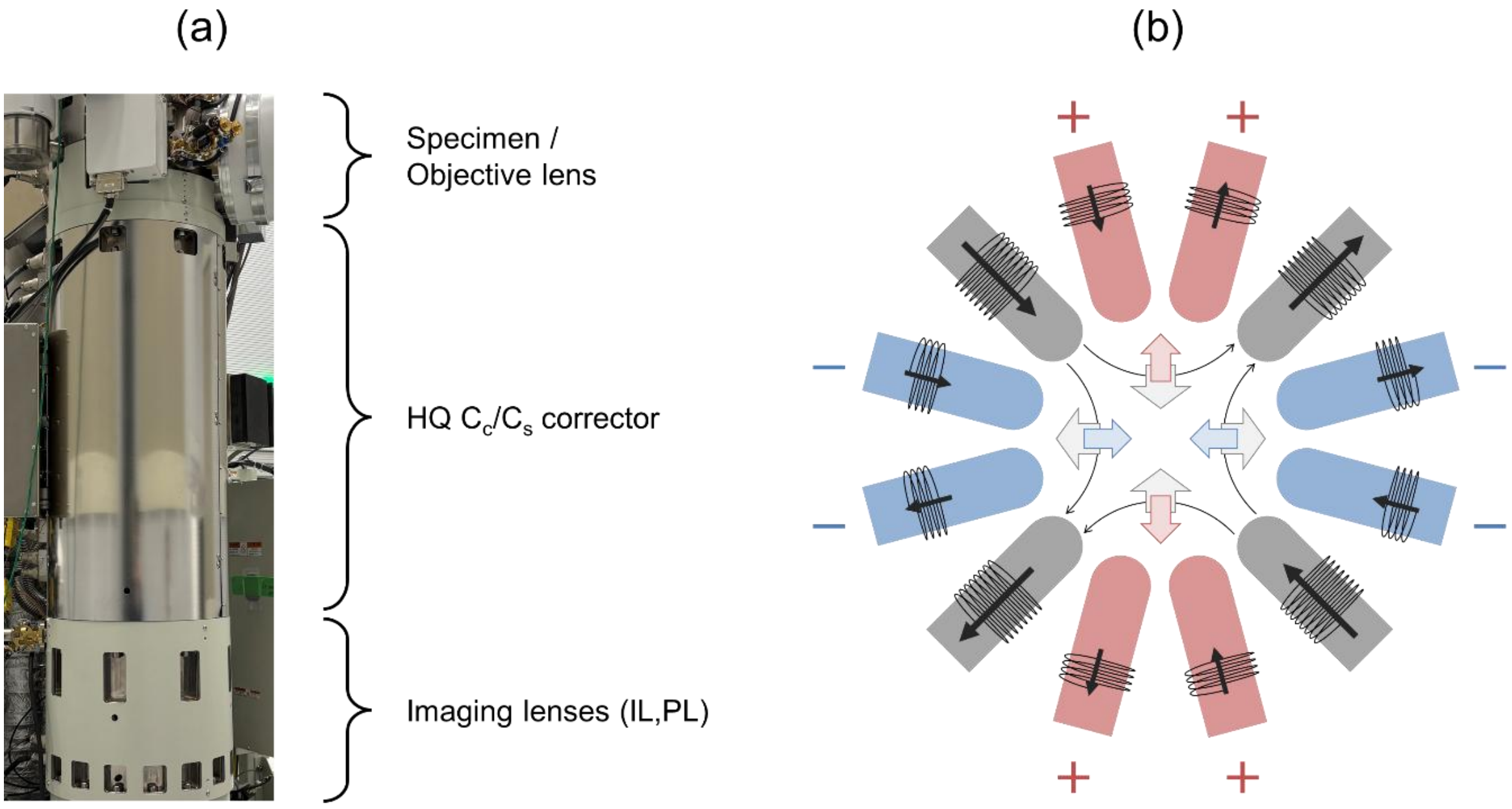

Figure 5: (a) The lower column of a transmission electron microscope with the HQ corrector installed between the objective and imaging lenses. (b) Schematic of a dodecapole used for the superposition of electric and magnetic quadrupole fields. Coils for other multipole fields are also wound around the individual poles.

The aberration coefficients $C_{c1,1,0}$ and $C_{c1,1,2}$ were evaluated by measuring the first-order axial aberrations from power spectra calculated from images recorded at different electron energies. For an instrument without correction, $C_c$ can be expressed as:

$$df(dE) = C_c \frac{dE^*}{E^*} = C_c \frac{1+eE/mc^2}{1+eE/2mc^2} \frac{dE}{E} \quad (19)$$

where $E$ and $dE$ are the primary electron energy and the energy deviation, respectively [31] and $E^*$ is the relativistic modified electric potential; $E^* = E(1 + eE/2mc^2)$. In the experiments the uncorrected value of $C_c$ of the imaging system was 1.1 mm, corresponding to 6.5 nm/eV. Following correction, both $C_{c1,1,0}$ and $C_{c1,1,2}$ were measured as both are first-order aberrations and $C_{c1,1,2}$ can be larger than $C_{c1,1,0}$. Following alignment of the instrument, the first-order chromatic aberrations were measured as:

$$\left(C_{c1,1,0} + C_{c1,1,2} \cos 2(\theta - \theta_0)\right) \alpha \frac{dE^*}{E^*} \qquad (20)$$

where $\alpha$, $\theta$, $\theta_0$ represent the angle, azimuth, and azimuthal direction of $C_{c1,1,2}$. Experimentally, both $C_{c1,1,0}$ and $C_{c1,1,2}$ were corrected to values $C_{c1,1,0} \approx C_{c1,1,2} <$ 0.01 mm, corresponding to less than 0.06 nm/eV. To measure the residual chromatic aberrations, the first order energy dependent defocus [13] and astigmatism were measured and fitted with a polynomial of the form:

$$C_{ci,n,m}(dE) = -\sum_i C_{ci,n,m} \left(\frac{dE^*}{E^*}\right)^i \qquad (21)$$

Fig. 6a shows that after the correction of the first order coefficients $C_{c1,1,0}$ and $C_{c1,1,2}$, the second order coefficients $C_{c2,1,0}$ and $C_{c2,1,2}$ are dominant and that the second order chromatic astigmatism is larger than the second order chromatic defocus *i.e.* $C_{c2,1,2} > C_{c2,1,0}$. Although these higher order coefficients remain uncorrected their effects are not significant if the first order axial chromatic aberrations are corrected.

In addition to the first order chromatic aberrations, axial geometric aberrations including $C_{3,0}$ were also corrected. For the measurement of the residual axial geometric aberrations, a conventional tableau of power spectra calculated from images recorded for a known set of incident illumination tilt angles [32] was used. Following correction of $C_c$ using quadrupoles, hexapole fields were excited to generate a negative $C_{3,0}$. The tableau shown in Fig. 6b shows correction of the original $C_{3,0}$ of the objective lens (0.7 mm). The results also demonstrate that the quadrupole multiplets serve the same function as a conventional round lens transfer doublet and the corrected value of $C_{3,0}$ is equivalent to that measured after correction using a conventional hexapole-type $C_{3,0}$ corrector. Importantly, higher order geometric axial aberrations with symmetries of 2 or 4, which are usually large in QO corrector are reduced in this corrector design because the strong octupole fields required for $C_{3,0}$ correction in QO correctors are not used in the HQ corrector, and high order coma and five-fold astigmatism are minimised due to the symmetric trajectories in the two quadrupole quadruplets. In addition, six-fold astigmatism, which is an intrinsic aberration in hexapole correctors, was corrected using weak octupole fields superimposed on the quadrupole fields. In these initial experiments, although remaining 5$^{th}$ order aberrations limit the aberration free angular range, it is possible to correct these by carefully optimising the alignment using the additional weak multipole fields included in the corrector. However, an aberration free angular range of ca. 35 mrad (half angle) has been obtained at 200 kV which is sufficient for a resolution better than 0.1 nm.

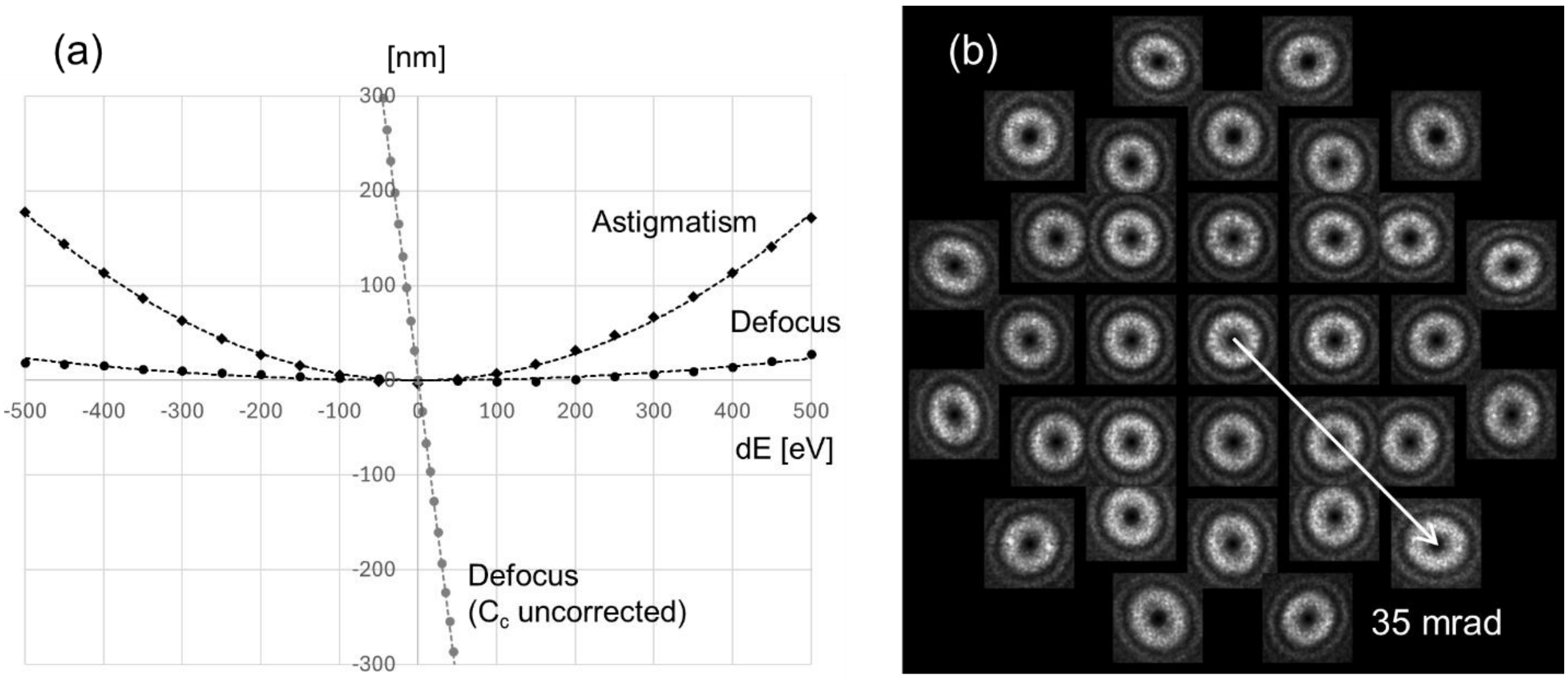

Figure 6: Experimental results demonstrating $C_c/C_s$ correction. (a) Energy-dependent defocus and astigmatism. The deviations from dE = 0 are referenced to zero defocus. (b) Tableau of power spectra for an outer illumination tilt magnitude of 35mrad (half angle).

**Images with $C_c/C_s$ Correction**

Following $C_c/C_s$ correction condition, TEM images of supported gold nanoparticles were recorded at 200 kV. To demonstrate the effect of $C_c/C_s$ correction with large energy spreads, an oscillation of the electron energy was intentionally induced during image recording. Figure 7 shows power spectra calculated from TEM images acquired at a defocus of -500 nm with an exposure time of 1 second with an energy oscillation amplitude of 100 eV at a frequency of 1.1 Hz. Without correction, the uncorrected value of $C_c$ =1.1 mm gives a defocus deviation of 640 nm for an energy deviation of 100 eV. Hence, in the power spectra calculated from images without correction, only blurred Thon rings from the amorphous carbon support film and a weak {111} reflection from the gold particles at 0.23 nm are visible. In contrast, after correction, Thon rings extending to high spatial frequency and the {220} reflection from the gold particles at 0.14 nm are clearly resolved.

We have also demonstrated $C_c/C_s$ correction using high-resolution TEM images of gold nanoparticles at a defocus of -20 nm with an exposure time of 1 second as shown in fig. 8. Even when an energy oscillation of ±100 eV at a frequency of 1.1 Hz was induced, atomic columns in the gold particles are visible. Although the resolution of the image with an oscillation of ±200 eV is lower than that without the energy oscillation as expected, atomic columns are still observed in some particles and reflections corresponding to the {220} spacing are present. These results suggest that high resolution images can be obtained even for electron sources with large energy spreads. In this regard the $C_c$ corrector maybe particularly useful for achieving high spatial resolution in time-resolved single-shot imaging employing short duration electron pulses at intermediate voltages with large beam currents where the Boersch effect [33] leads to a significant energy broadening.

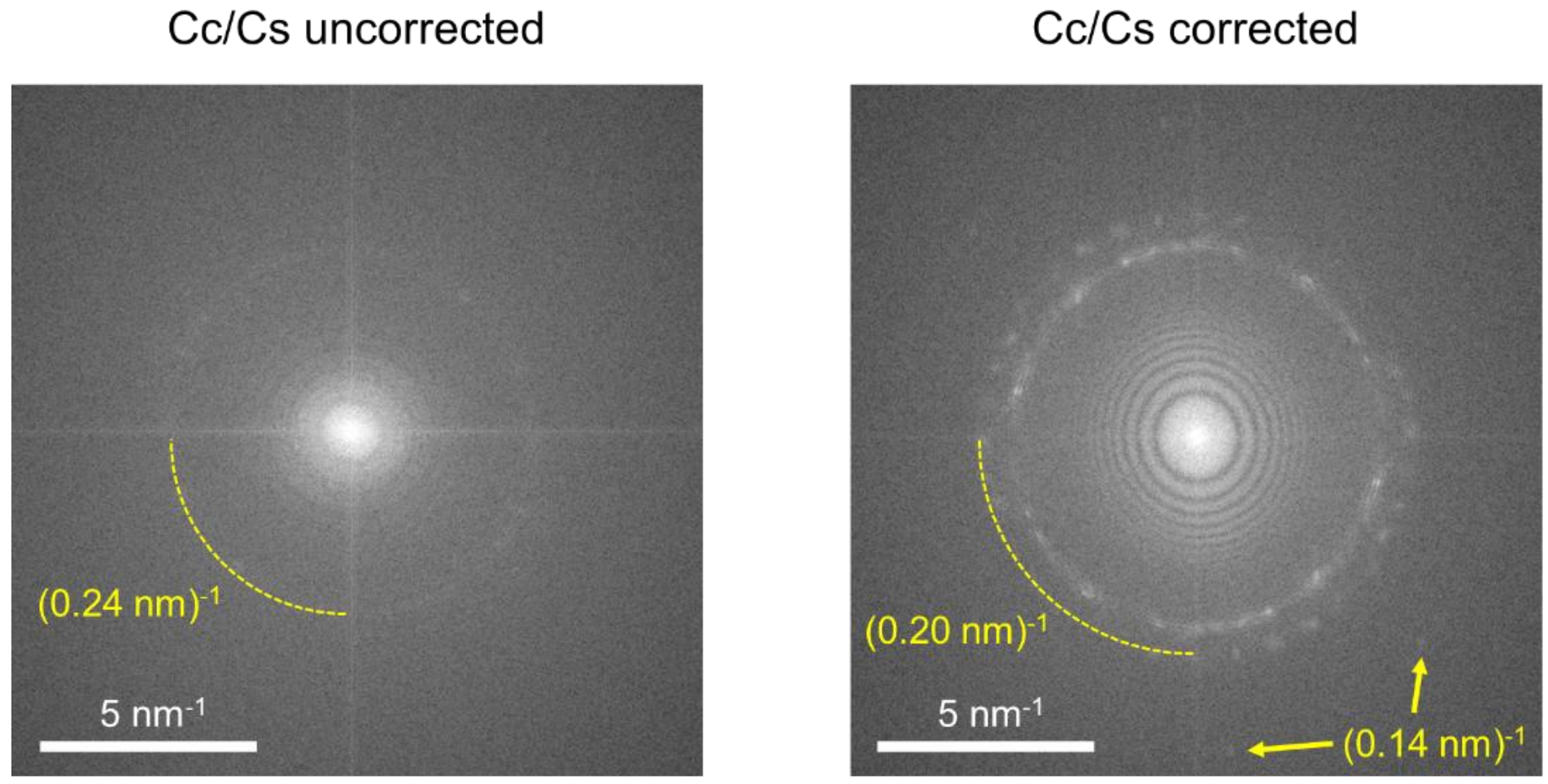

Figure 7: Power spectra calculated from TEM images of gold nanoparticles with and without $C_c/C_s$ correction. Images were taken with an oscillation of the beam energy during recording with an amplitude of 100 eV.

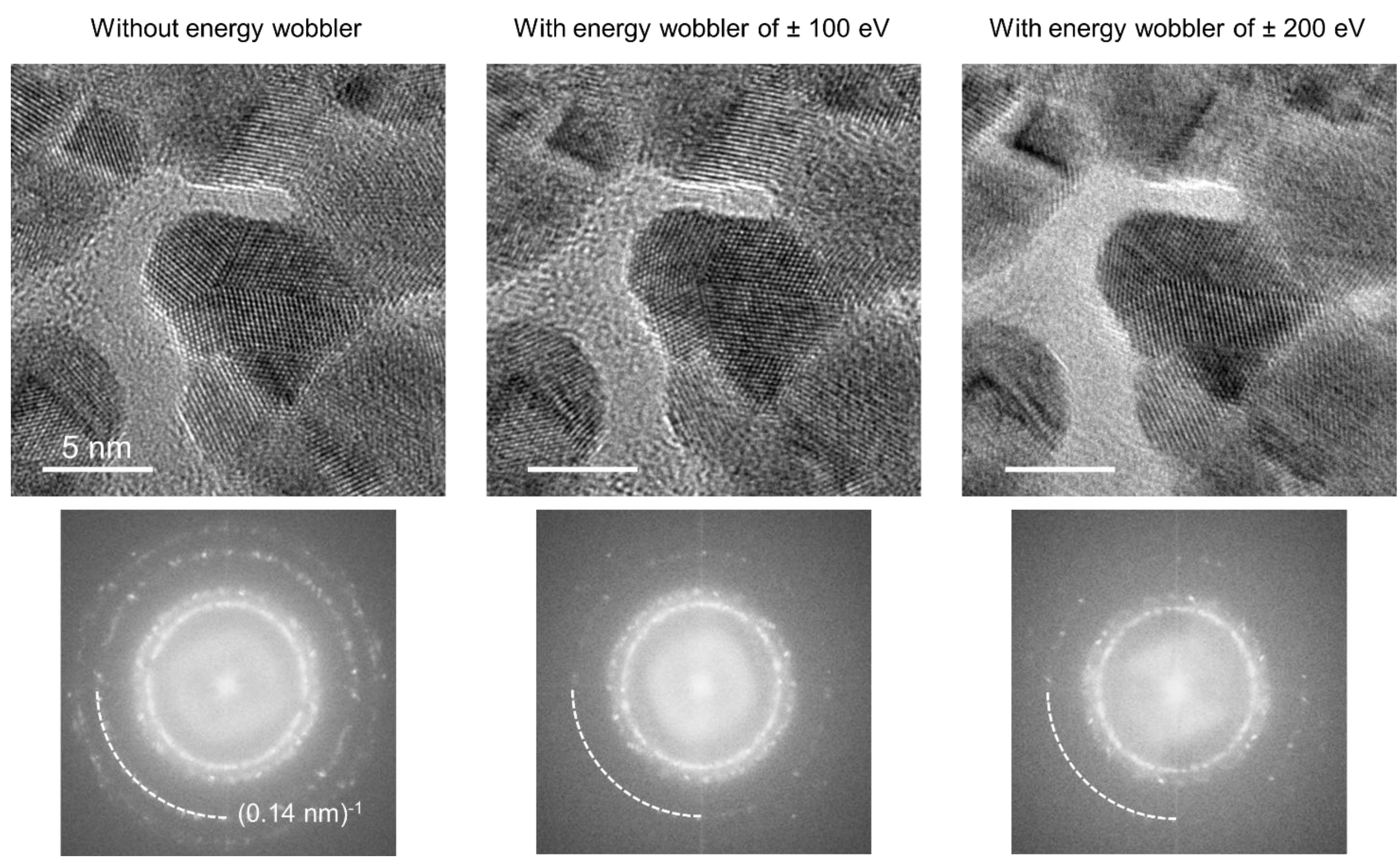

Figure 8: TEM images of gold nanoparticles following $C_c/C_s$ correction condition with various magnitude electron energy wobblers. The power spectra calculated from each image in the upper row are shown in the lower row.

## Summary and Conclusions

We have reported a novel design for a hexapole-quadrupole $C_c/C_s$ corrector for a transmission electron microscope and have demonstrated correction of first order chromatic aberrations and geometric axial aberrations to third order. Importantly, correction of the geometric aberrations including $C_{3,0}$ shows that the quadrupole multiplet inherent to this design can be used instead of a conventional round lens transfer doublets. Furthermore, the required hexapole elements can be placed at conjugate planes and thus generation of undesirable combination aberrations can be avoided. We have illustrated correction of $C_c$ using TEM images with an intentionally induced energy oscillation in which spacings < 0.14 nm can be resolved even with an effective energy spread > 100 eV. This suggests that the atomic resolution images can be recorded even using an electron source with a large energy spread with potential application to time-resolved single-shot imaging with high-density electron pulses.

## Acknowledgements

Funding from the Engineering and Physical Sciences Research Council to the Rosalind Franklin Institute is acknowledged. The authors thank O Krivanek and N Dellby for helpful discussions on the appropriate general form of their reported aberration nomenclature [3] to include chromatic terms.